\documentstyle[aps,12pt,psfig]{revtex}

\newcommand{ \be }{\begin{equation}}
\newcommand{ \ee }{\end{equation}}
\newcommand{ \bea }{\begin{eqnarray}}
\newcommand{ \eea }{\end{eqnarray}}
\newcommand{ \la }{\langle}
\newcommand{ \ra }{\rangle}
\newcommand{ \bp }{{\bf p}}
\newcommand{ \bQ }{{\bf Q}}

\begin{document}

\title{Methods for analyzing anisotropic flow in relativistic nuclear collisions} 

\author{
A.~M.~Poskanzer$^{1)}$
and
S.~A.~Voloshin$^{2)}$\footnote{On leave from Moscow 
Engineering Physics Institute, Moscow, 115409, Russia} 
}

\address{
1) Nuclear Science Division, Lawrence Berkeley National Laboratory,
Berkeley, CA \\ 
2) Physikalisches Institut der Universit\"{a}t
Heidelberg, Germany }

\date{\today}
\maketitle

\begin{abstract} 
The strategy and techniques for analyzing anisotropic flow (directed,
elliptic, etc.)  in relativistic nuclear collisions are presented.
The emphasis is on the use of the Fourier expansion of azimuthal
distributions.  We present formulae relevant for this approach, and in
particular, show how the event multiplicity enters into the event
plane resolution.  We also discuss the role of non-flow correlations
and a method for introducing flow into a simulation.
\end{abstract}

\centerline{Phys. Rev. C  no. CS6346}
\centerline{Received 20 May 98}
\centerline{Draft 3.9}

\narrowtext

\section{Introduction: Subject and terminology}
Recently, the study of collective flow in nuclear collisions at high
energies has attracted increased attention of both theoreticians and
experimentalists~\cite{olli98}.  There are several reasons for this:
i) the observation of anisotropic flow at the
AGS~\cite{l877flow1,l877flow2,l877flow3,ahle98,volo98,liu98} and at
the SPS~\cite{lna49,posk98,lna45,lwa98}, ii) progress in the
theoretical understanding of the relation between the appearance and
development of flow during the collision evolution, and processes such
as thermalization, creation of the quark-gluon plasma, phase
transitions, etc.~\cite{ljs,lbrav,lko,lshur,risc96,dan95,dan98}, iii)
the study of mean field effects~\cite{lko,lsorge}, iv) the importance
of flow for other measurements such as identical and non-identical
two-particle correlation analyses~\cite{lcsorgo,lvc,wied97,volo97}, v)
the development of new techniques suitable for flow studies at high
energies~\cite{olli92,lvzh,ollimeth,olli98,posk98}.  Although all
forms of flow are interrelated and represent only different parts of
one global picture, usually people discuss different forms of
collective flow, such as longitudinal expansion, radial transverse
flow, and anisotropic transverse flow of which the most well
established are directed flow and elliptic~\cite{olli92,lsorge}
flow~\cite{H-G}.

The study of flow in ultrarelativistic nuclear collisions has just
begun and it is very important to define the subject and establish the
terminology.  At high energies the longitudinal flow is well decoupled
from transverse flow.  The flow (polar) angles observed at low
energies are relatively large and a rotation of the coordinate system
was done in order to analyze the event shape in the plane
perpendicular to the main axis of the flow ellipsoid. At high energies
the flow angles become very small, $\theta_{flow} \approx \la p_x
\ra/p_{beam} \ll 1$, so that not doing such a rotation ``induces'' an
elliptic anisotropy of the order of the square of the flow angle and
usually can be neglected. This means that at high energies one does
not have to rotate to the flow axis to study the flow
pattern~\cite{H-G}, but one can use the plane transverse to the beam
axis.  Thus we discuss only anisotropic transverse flow from the
particle azimuthal distributions at fixed rapidity or pseudorapidity.
It appears very convenient to describe the azimuthal distributions by
means of a Fourier expansion~\cite{Bev,lvzh,olli92}, and below we
characterize different kinds of anisotropies as corresponding to
different harmonics (in analogy/contrast to the description at low
energies, where the three-dimensional event shape is characterized
using multipole terminology).  Anisotropic flow corresponding to the
first two harmonics plays a very important role and we use special
terms for them: directed and elliptic flow, respectively. The word
``directed'' comes from the fact that such flow has a direction, and
the word ``elliptic'' is due to the fact that in polar coordinates the
azimuthal distribution with non zero second harmonic represents an
ellipse.

\section{Correlations with respect to an event plane}
{\bf Strategy}. 
In this section we summarize an approach to the study of anisotropic
flow which is particularly suitable at high (AGS/SPS/RHIC) energies.
It uses the Fourier expansion of azimuthal distributions, introduced
in this way for the analysis in ref.~\cite{lvzh}.  The essence of the
method is to first estimate the reaction plane.  The estimated
reaction plane we call the event plane.  The Fourier coefficients in
the expansion of the azimuthal distribution of particles with respect
to this event plane are evaluated.  Because the finite number of
detected particles produces limited resolution in the angle of the
measured event plane, these coefficients must be corrected up to what
they would be relative to the real reaction plane.  This is done by
dividing the observed coefficients by the event plane resolution,
which is estimated from the correlation of the planes of independent
sub-events~\cite{dan85} (sub-groups of the particles used for the
event plane determination). The resolution obtained from the
sub-events can be converted to that for the full event by means of the
multiplicity dependence of the resolution which will be described
below.  Also, if the detector does not have full azimuthal acceptance,
the acceptance bias has to be removed.

{\bf Fourier expansion}. 
The quantity under study in the most general case is the triple
differential distribution. In this, the dependence on the particle
emission azimuthal angle measured with respect to the reaction plane
can be written in a form of Fourier series 
\be
 E \frac{d^3 N}{d^3 p} = 
 \frac{1}{2\pi} \frac {d^2N}{p_t d p_t dy}(1+
 \sum_{n=1}^\infty 2 v_n \cos(n(\phi-\Psi_r))),
\label{ed3t}
\ee
where $\Psi_r$ denotes the (true) reaction plane angle, and the sine
terms vanish due to the reflection symmetry with respect to the
reaction plane.  The main advantage of the Fourier method is that the
Fourier coefficients, evaluated using observed event planes, can be
corrected for the event plane resolution caused by the finite
multiplicity of the events. This correction always raises the value of
the coefficients. The great importance of this is that then the
results for particles in a certain phase space region may be compared
directly to theoretical predictions, or to simulations unfiltered for
the detector acceptance, and for which the reaction plane has been
taken to be the plane containing the theoretical impact parameter.
Note the factor of ``2'' in front of each $v_n$ coefficient.  We
propose to use it because in this case the meaning of the coefficients
$v_n$ becomes transparent~\cite{lvzh}, $v_n=\la \cos(n(\phi-\Psi_r))
\ra$, where $\la \: \ra$ indicates an average over all particles in
all events.  For the particle number distribution, the coefficient
$v_1$ is $\la p_{x} / p_{t} \ra$ and $v_2$ is $\la ( p_x / p_t )^{2} -
( p_y / p_t )^{2}
\ra$.

{\bf Estimation of the reaction plane}.  
The method uses the anisotropic flow itself to determine the event
plane.  It also means that the event plane can be determined
independently for each harmonic of the anisotropic flow.  The event
flow vector $Q_n$ and the event plane angle $\Psi_n$ from the $n-$th
harmonic of the distribution are defined by the equations
\bea
 Q_n \cos(n\Psi_n) &=& X_n= \sum_i w_i \cos (n\phi_{i}),
\label{x}
\\
 Q_n \sin(n\Psi_n) &=& Y_n= \sum_i w_i \sin (n\phi_{i}),
\label{y}
\eea
or
\be
 \Psi_n = \left( \tan^{-1} 
 \frac{\sum_i w_i \sin (n\phi_{i})}{\sum_i w_i \cos (n\phi_{i})}
 \right) / n.
\label{arctan}
\ee
The sums go over the $i$ particles used in the event plane
determination and the $w_i$ are weights. In general the weights are
also optimized to make the reaction plane resolution the best that is
possible.  Sometimes it can be done by selecting the particles of one
particular type, or weighting with transverse momentum of the
particles, etc.  Usually the weights for the odd and even harmonic
planes are different. Optimal weights are discussed in footnote 2
of~\cite{dan95}. For symmetric collisions reflection symmetry says
that particle distributions in the backward hemisphere of the center
of mass should be the same as in the forward hemisphere if the
azimuthal angles of all particles are shifted by $\pi$.  This explains
why for the odd harmonics the signs of the weights are reversed in the
backward hemisphere while for the even harmonics the signs of the
weights are not reversed.  Note that the event plane angle $\Psi_n$
determined from the $n-$th harmonic is in the range $0
\le \Psi_n < 2\pi/n$.  For the case of $n=1$,
Eqs.~(\ref{x},\ref{y},\ref{arctan}) are equivalent to obtaining
$\Psi_1$ for number flow from\cite{dan85}
\be
 {\bf Q}  = \sum{w \bp_t / |p_t|}
\label{eqv}
\ee
where the sum is over all the particles. The case of $n=2$ is
equivalent to the event plane determined from the transverse
sphericity matrix~\cite{olli92}.

{\bf Acceptance correlations}.
Biases due to the finite acceptance of the detector which cause the
particles to be azimuthally anisotropic in the laboratory system can
be removed by making the distribution of event planes isotropic in the
laboratory. We know of several different methods to remove the effects
of anisotropy which have been used (sometimes in a combination with
each other).  Each of them has some advantages along with
disadvantages.

The simplest one is to
re-center~\cite{l877flow2,l877flow3,lna45,dan88} the distributions
$(X_n,Y_n)$ (Eqs.~\ref{x},\ref{y}) by subtracting the $(X_n,Y_n)$
values averaged over all events.  The main disadvantage of this method
is that it does not remove higher harmonics from the resulting
distribution of $\Psi_n$. If such harmonics are present then the
method requires additional flattening of the event plane distribution
by one of the other methods.  The second, one of the most commonly
used methods, is to use the distribution of the particles themselves
as a measure of the acceptance~\cite{l877flow2,l877flow3,posk98}. One
accumulates the laboratory azimuthal distribution of the particles for
all events and uses the inverse of this as weights in the above
calculation of the event planes.  The limitation of this approach is
that it does not take into account the multiplicity fluctuations
around the mean value.  The third method is to use mixed
events~\cite{l877flow2,l877flow3,posk98}.  Correlations with the raw
event planes are stored in histograms and correlations with event
planes from previous events are also stored. The real correlations are
then divided by the mixed event correlations to remove the acceptance
correlations. This third method suffers from the problem that, if one
uses only one mixed event for each real event, the errors are
$\sqrt{2}$ larger. If one uses many mixed events for each real event,
the errors decrease as $n_{mix}^{1/4}$ instead of $\sqrt{n_{mix}}$
because the same events are being used $n_{mix}$ times~\cite{lvmix}.
The fourth method fits the unweighted laboratory distribution of the
event planes, summed over all events, to a Fourier expansion and
devises an event-by-event shifting of the planes needed to make the
final distribution isotropic~\cite{l877flow2,l877flow3}\footnote{The
equation for the shift is (see Appendix in~\cite{l877flow2})
\be
n \Delta \Psi_n = \sum_{i=1}^{i_{max}} \frac{2}{i} 
( - \la \sin (i n \Psi_n ) \ra \cos (i n \Psi_n )
+ \la \cos (i n \Psi_n ) \ra \sin (i n \Psi_n ) ) .
\ee
We have usually taken $i_{max}=4/n$ for $n=1,2$.}.  In all these
methods one has to check that the event plane distributions indeed
become isotropic.

{\bf Particle distributions with respect to the event plane}.
The next step is to study the particle distributions with respect to
the event planes.  Note that for a given $n$ the corresponding Fourier
coefficient $v_n$ can be evaluated using the reaction planes
determined from any harmonic $m$, with $n \geq m$, if $n$ is a
multiple of $m$.  If $n > m$, the sign of $v_n$ is determined relative
to $v_m$.  That is, the first harmonic plane can be used, in
principle, to evaluate all $v_n$.  The second harmonic plane can be
used to evaluate $v_2$, $v_4$, etc.  For the event plane evaluated
from the $m$-th harmonic the Fourier expansion is
\be
 \frac{d(w N)}{d(\phi-\Psi_m)}=\frac{\la w N \ra}{2\pi}
 (1+
\sum_{k=1}^\infty 2v_{k m}^{obs}\cos(km(\phi-\Psi_m))).
\label{vn}
\ee
Writing the equation in terms of $km$ instead of $n$ insures, for
instance, that when $m = 1$ all terms are present, but when $m = 2$
only the even terms are present. The quantity $w$ is a weight which
could be $p_t$ of the particle if one studies transverse momentum
flow, or just unity, if one studies flow of the number of particles.
The coefficients $v_n^{obs}$ are evaluated by
$\la\cos(n(\phi-\Psi_m))\ra$. The quantity $(\phi-\Psi_m)$ has a
lowest order periodicity of $2\pi/m$. To graph the distribution one
can shift the negative values to positive by adding $2
\pi$, and then fold the distribution with this periodicity using the
module function. When a particle has been used in the calculation of
an event plane, the auto-correlation effect in its distribution with
respect to this plane is removed by recalculating that plane without
this particle~\cite{dan85}. This is easily done if one saves the sums
of sines and cosines from Eqs.~(\ref{x}-\ref{arctan}), and subtracts
the contribution of the particle and its weight from these sums. This
method of removing autocorrelations assumes that contributions from
conservation of momentum are small.

{\bf The event plane resolution}.
The coefficients in the Fourier expansion of the azimuthal
distributions with respect to the {\em real} reaction plane are then
evaluated by dividing by the event plane
resolution~\cite{dan85,demoul,lvzh,l877flow2,l877flow3,ollimeth},
\be
 v_n=v_n^{obs}/\la \cos(k m (\Psi_m-\Psi_r) \ra.
\ee 
The mean cosine values are less than one and thus this correction
always increases the flow coefficients.

The resolution depends both on the harmonic used to determine the
event plane $m$ and the order of the calculated coefficient $n$.  It
is generally true that better accuracy for the determination of $v_n$
is achieved by using the event plane ($\Psi_n$) determined from the
same harmonic ($m=n$, $k=1$) because the resolution deteriorates as
$k$ increases (see below).  For example, better accuracy for $v_2$ can
be achieved using $\Psi_2$ even when the elliptic flow is somewhat
weaker than the directed flow.

To calculate the resolution we start with the distribution of
$m(\Psi_m-\Psi_r)$, which can be written as~\cite{lvzh}
\be
 \frac{dP}{d(m(\Psi_m-\Psi_r))}
 =\int \frac{v_m' dv_m' }{2\pi \sigma^2}
 \exp(-\frac{v_m^2+v_m'^2-2v_m v_m' \cos(m(\Psi_m-\Psi_r))} {2 \sigma^2}).
\label{integral}
\ee
The parameter $\sigma$, which to second order in flow is common for all
$m$, is inversely proportional to the square-root of $N$, 
the number of particles used to determine the event plane
\be
 \sigma^2=\frac{1}{2N} \frac{\la w^2 \ra}{\la w \ra^2}.
\label{sigma}
\ee

The integral (\ref{integral}) can be evaluated
analytically~\cite{olli92,lvzh}, and then the event plane resolution
can be expressed as
\be
 \langle \cos (km(\Psi_m-\Psi_r)) \rangle 
=\frac{\sqrt{\pi}}{2\sqrt{2}}
 \chi_m \exp (-\chi_m^{2}/4)
 \left[ I_{\frac{k-1}{2}} (\chi_m^{2} / 4) +
 I_{\frac{k+1}{2}} (\chi_m^{2} / 4) \right],
\label{emeancos}
\ee
where $\chi_m \equiv v_m/\sigma$ ($=v_m\sqrt{2N}$ for number
flow)\footnote{Please note that the parameter $\chi$ used in this
paper is a factor of $\sqrt{2}$ larger than the one used in
refs.~\cite{olli98,ollimeth}; it is equivalent to the parameter
$\tilde{\xi}$ in~\cite{lvzh}.}  and $I_\nu$ is the modified Bessel
function of order $\nu$.  This resolution function is plotted in
Fig. 1.

Note that $\langle \cos (m(\Psi_m-\Psi_r)) \rangle$ is a correction
for the reconstruction of the signal if the event plane is derived
using the flow of the same harmonic ($k=1$).  In the case when the
harmonic orders do not coincide, for example, when one uses the event
plane derived from the first harmonic (directed flow) and studies the
second harmonic (elliptic flow) of the particle distribution with
respect to this plane (the $m=1, k=2$ term in Eq.~(\ref{vn})) the
correction would be $\langle \cos (2(\Psi_1-\Psi_r)) \rangle$.  For
practical use all such functions can be calculated numerically as in
the section on {\it Approximations} below.

The resolution correction used in earlier times at the
Bevalac~\cite{Bev} is close to the present curve for values of
$\chi_1=v_1\sqrt{2 N}$ greater than about 1.5, which was generally
true at the Bevalac. However, using the old procedure at the higher
beam energies of the AGS or SPS would greatly overestimate the
resolution and make the flow values too small.

{\bf The correlation between flow angles of independent sets of
particles}.
If one constructs the event planes in two different windows, (a) and
(b), or from two random sub-events\footnote{The random sub-events can
be made to have equal multiplicity using the CERN library routine
SORTZV.}, the corresponding correlation function also can be written
analytically.  But more important in this case is the simple relation
for such correlations,
\be
 \la \cos (n(\Psi_n^{a}-\Psi_n^{b})) \ra
 = 
 \la \cos (n(\Psi_n^{a}-\Psi_r)) \ra
 \la \cos(n(\Psi_n^{b}-\Psi_r)) \ra.
\label{sub-corr}
\ee
Here, the assumption is made that there are no other correlations
except the ones due to flow, or that such other correlations can be
neglected.  (For example, two-particle correlations due to resonance
decays should scale with multiplicity as $1/N$ and usually can be neglected.)
If this is not true, then special precautions have to be made to avoid
or correct for such correlations. See the section below on {\it Non-flow
Correlations}.

Note that the correlation of two planes (the distribution in
$\Psi^a-\Psi^{b}$) is {\it not} well represented by a Fourier
expansion. For a strong correlation it should be Gaussian. For no flow
the correlation between two random planes of the same order should be
a triangular distribution. For plotting purposes, if one takes the
absolute value of $\Psi_n^{a}-\Psi_n^{b}$ and folds this distribution
about the angle $\pi/n$, the triangle becomes flat. This does not
affect the $\la \cos (n(\Psi_n^{a}-\Psi_n^{b})) \ra$ value needed for
Eq.~\ref{sub-corr}, but makes it easier to see a real flow effect by
the non-flatness of the distribution in the graph.

For the correlation between angles determined from two sub-events of
different harmonics one can write similar relations. For example, the
correlation between $\Psi_1^{a}$ and $\Psi_2^{b}$ is
\be
 \la \cos (2(\Psi_1^{a}-\Psi_2^{b})) \ra
 = 
 \la \cos (2(\Psi_1^{a}-\Psi_r)) \ra
 \la \cos(2(\Psi_2^{b}-\Psi_r)) \ra
\ee
This expression is proportional to $v_1^2 v_2/\sigma^3$ and can be
rather small in magnitude, but it can be useful for the determination
of the relative orientation of the flow of different orders (see also
the section on {\it Approximations} below).  On the other hand, the
relative orientation also can be determined using Eq.~(\ref{vn}) with
$k > 1$.

\begin{figure}
\centerline{\psfig{figure=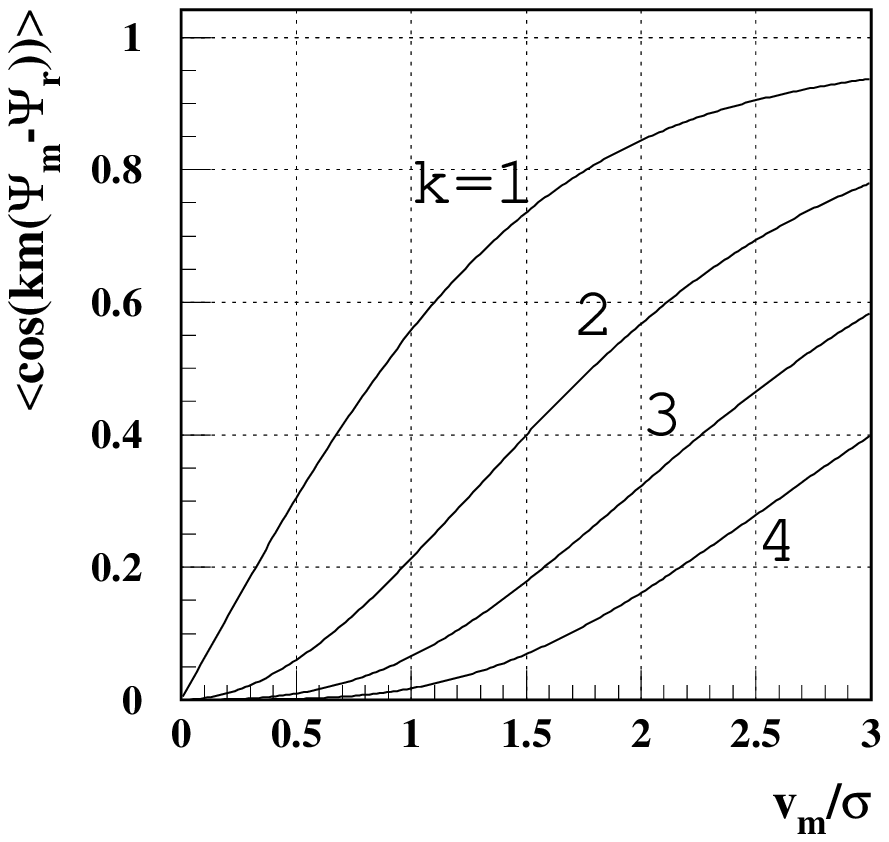,height=12.0cm}}
  \caption[]{ The event plane resolution for the $n-$th ($n=km$)
harmonic of the particle distribution with respect to the $m-$th
harmonic plane, as a function of $\chi_m=v_m/\sigma$.}
\label{f1}
\end{figure}

{\bf The determination of the event plane resolution}.
The above relations permit the evaluation of the event plane
resolution directly from the data.  For example, if one knows the
correlation between two equal multiplicity sub-events (where the
resolution of each is expected to be the same) then from
Eq.~(\ref{sub-corr}) the resolution of each of them is
\be
 \la \cos(n(\Psi_m^{a}-\Psi_r)) \ra
 = \sqrt{\la \cos(n(\Psi_m^{a}-\Psi_m^{b})) \ra},
\label{mcos}
\ee
where, as before, $n=km$, and $k$ is not necessarily equal to 1. If
the sub-events are correlated, then the term inside the square-root is
always positive.\footnote{For small amounts of flow, fluctuations can
cause this term to be negative for even harmonics when the total $p_t$
is required to be zero.} Note that the event plane resolution
determined in such a way is the event plane resolution of the
sub-events.  If one wants to use the full event (all detected
particles from the event) to determine the event plane, then the full
event plane resolution can be calculated from the sub-event resolution
using Eq.~(\ref{emeancos}) or the approximations below,
Eqs.~(\ref{rescorr1},\ref{rescorr21}), taking into account that the
multiplicity of the full event is twice as large as the multiplicity
of the sub-event.  Because $\chi_m=v_m/\sigma$ is proportional to
$\sqrt{N}$, in a case of low resolution where the curves in
Fig.~\ref{f1} are linear, this reduces to
\be
\la \cos(n(\Psi_m-\Psi_r)) \ra =
\sqrt{2} \la \cos(n(\Psi_m^{a}-\Psi_r)) \ra.
\ee
If the sub-events are not ``equal'', or if one has only correlations
between particles in different windows, and the resolution in each
window can be different, then one needs at least three windows to
determine the event plane resolution in each of them.  In this case,
for example, the resolution in the first window is determined
as~\cite{l877flow2,l877flow3}
\be
 \la \cos(n(\Psi_m^{a}-\Psi_r)) \ra
 = \sqrt{ \frac{\la \cos(n(\Psi_m^{a}-\Psi_m^{b})) \ra
 \la \cos(n(\Psi_m^{a}-\Psi_m^{c})) \ra } 
 {\la \cos(n(\Psi_m^{b}-\Psi_m^{c})) \ra}}.
\ee

{\bf Approximations}.
We want to discuss two limits.  In a case of weak flow, $\chi_m \ll 1$
($v_m \ll \sigma$, which also means that the event plane resolution is
low) one can expand the exponent under the integral in
Eq.~(\ref{integral}) before the integration.  It will follow that in
this limit
\be
 \frac{dP}{d(m(\Psi_m-\Psi_r))}
 \propto 1+ \sqrt{\frac{\pi}{2}} \chi_m \cos(m(\Psi_m-\Psi_r)).
\ee
Then the reaction plane resolution can be estimated analytically to be
(the case $k=1$)
\be
 \la \cos(m(\Psi_m-\Psi_r)) \ra \approx \sqrt{\frac{\pi}{8} }\chi_m.
\label{eqcos1}
\ee
In this case the resolution is linear in $\chi$ and, since $\sigma$ is
the same for all $m$, the resolution for different $m$ scales with
$v_m$.

Combining Eqs.~(\ref{mcos},\ref{sigma},\ref{eqcos1}) one can estimate
the flow signal directly from the correlation between two sub-events
\be
 v_m \approx \sqrt{\frac{4}{\pi} \frac{\la w^2 \ra}{\la w \ra^2}} \sqrt{ 
 \frac{ \la \cos(m(\Psi_m^{a}-\Psi_m^{b})) \ra } {N_{sub}} },
\ee
where $N_{sub}$ is  the multiplicity of the sub-events.

Keeping the second order terms in the expansion of
Eq.~(\ref{integral}) it can be shown that
\be
 \la \cos(2m (\Psi_m-\Psi_r)) \ra \approx \frac{\chi_m^2}{4} 
 \approx \frac{2}{\pi} 
 \left(\la \cos(m(\Psi_m-\Psi_r)) \ra \right) ^2.
\label{eqcos2}
\ee
The latter relation for $k=2$ is needed as a resolution correction for
$v_2$ in the case where the event plane was determined by the directed
flow ($m = 1$).  The approximate relations (\ref{eqcos1},\ref{eqcos2})
are rather accurate for $\chi_m<0.5$.

Using the above approximations the correlation between flow angles of
sub-events of different harmonics ($m=1$ and $m=2$) can be written as
\be
 \la \cos (2(\Psi_1^{a}-\Psi_2^{b})) \ra 
\approx 
\pm \frac{2}{\pi} \la \cos (\Psi_1^{a}-\Psi_1^{b}) \ra
\sqrt{\la \cos(2(\Psi_2^{a}-\Psi_2^{b})) \ra}.
\ee
As stated above, the sign of the left hand side shows the relative
orientation of the flow of the different harmonics, while the right
hand side after the $\pm$ sign is always positive. Note that all
quantities in this equation are evaluated from the data independently,
so the comparison of the magnitudes is an additional consistency check.

In the second limiting case of strong flow $\chi_m \gg 1$, one can
expand the cosine in the exponent of Eq.~(\ref{integral}) and get
approximately
\be
 \frac{dP} {d(m(\Psi_m-\Psi_r))}
 \approx 
\frac{\chi_m}{\sqrt{2\pi}}
\exp{(- \frac{\chi_m^2 (m(\Psi_m-\Psi_r)^2)}{2})},
\ee
which also can be used to calculate the reaction plane resolution.
The case of strong flow is important mostly for relatively low energy
collisions and we will not discuss it in detail.  Note that in this
case (and {\it only} in this case) the distribution of
$m(\Psi_m-\Psi_r)$ can be described by a Gaussian.

Here we present interpolation formulae of Eq.~(\ref{emeancos}) for 
the most needed cases of $k=1,2$
\be
  \la \cos (m(\Psi_m-\Psi_r)) \ra =
      0.626657 \chi_m   - 0.09694 \chi_m^3
     + 0.02754 \chi_m^4 - 0.002283 \chi_m^5,
\label{rescorr1}
\ee
\be
  \la \cos (2m(\Psi_m-\Psi_r)) \ra =
      0.25 \chi_m^2   - 0.011414 \chi_m^3
     - 0.034726 \chi_m^4 + 0.006815 \chi_m^5.
\label{rescorr21}
\ee
The interpolation formulae are valid for $\chi_m<3$.

In order to use the above equations to go from the sub-event
resolution to the full-event resolution, one has to set the equation
equal to the sub-event resolution and take the root to obtain
$\chi_m$. One multiplies this by $\sqrt{2}$ because $\chi_m$ is
proportional to $\sqrt{N}$, and then evaluates the full-event
resolution from the same equation. Our routines for finding the roots
of these equations use an iterative method, and the routines for
propagating the errors involve a calculation for the change in the
result with a small finite change in the input.

Another approximate method is to evaluate $\chi_m = v_m/\sigma$ of the
full event from the fraction of events where the correlation of the
planes of the sub-events is greater than $\pi/2$~\cite{ollimeth},
\be
\frac{ N_{events} (m | \Psi^{a}_m - \Psi^{b}_m | \: > \: \pi/2) }
{ N_{total} }
=
 \frac{e^{-\chi_m^{2}/4} }{ 2} .
\ee
This fraction, and therefore the equation, is only accurate when
$\chi_m$ is small.

\section{Event-by-event analysis of azimuthal distributions} 
\label{sebe}
Here we study the anisotropies in the azimuthal distributions of
particles within a relatively large rapidity (pseudorapidity) window
without determining an event plane.  Finite multiplicities used in
each event for the evaluation of the event flow vector ${\bf Q}_n$,
defined in Eqs.~(\ref{x},\ref{y}), causes both finite event plane
resolution and also fluctuations in the vector magnitude $Q_n$.  In
the case of zero flow $\la Q_n^2 \ra = \la N \ra $.  (In this section
for simplicity we assume $w_i=1$.)  Anisotropic flow, which shifts the
vector ${\bf Q}_n$ in each event in the (random) flow direction by a
value $v_{n}N$, results effectively in broadening the distribution in
$Q_n$.  Keeping the first non-vanishing contribution from flow one has
\be
\la Q_n^2 \ra = \la N \ra + \bar{v}_n^2 \la N^2 \ra.
\label{r2}
\ee
This equation can be used to estimate the flow signal $\bar{v}_n$, 
which is the average of $v_n$ over the rapidity region used for 
the $Q_n$ calculation.

Another method~\cite{lvzh} involves fitting the distribution in
$r_n=Q_n/N$ to the theoretical formula, which has $\bar{v}_n$ (more exactly
its absolute value) as a parameter.  In this method the distribution
of $r_n$ has to be fitted by the function
\be
 \frac{dP}{r_n\,dr_n}  =  \frac{1}{\sigma^2}
 \exp(-\frac{\bar{v}_n^2+r_n^2}{2\sigma^2}) 
 I_0(\frac{r_n \bar{v}_n}{\sigma^2}),
\label{ebess}
\ee
where $I_0$ is the modified Bessel function, $\bar{v}_n$ is the
parameter of interest, and $\sigma$ is the parameter related to finite
multiplicity fluctuations.  The distribution (\ref{ebess}) was derived
using the central limit theorem, requiring the particle multiplicity
$N$ to be large.  Note that in principle Eq.~(\ref{r2}) does not have
this limitation, although effectively the ``signal to noise ratio'' is
proportional to $N$ and one would need $N$ large to apply the equation
to data.

Note also that both methods do not require the determination of the
event plane and can be performed within one (pseudo)rapidity window,
but they do require this window to be rather wide in order to have a
relatively large number of particles. The distribution (\ref{ebess})
can be used to select events with larger or smaller values of flow.

\section{Non-flow correlation contribution}
The methods described so far in this paper are correct when
correlations induced by flow dominate all others.  By all others we
mean correlations due to momentum conservation~\cite{dan85}, long and
short range two- and many-particle correlations (due to quantum
statistics, resonances, mini and real jet production, etc.).  Below,
we discuss the contribution of such non-flow correlations to the
correlation of the event planes determined from two independent
sub-events.  Very often the contribution of non-flow correlations
scales as $1/N$, where $N$ is the multiplicity of particles used to
determine the event plane.  But one should remember that the
contribution due to momentum conservation increases with the fraction
of particles detected, and that the relative contribution of
Bose-Einstein correlations can be independent of multiplicity (the
later can be important if the sub-events are formed by random division
of all particle from the same phase space region into two subgroups,
where the particles contributing to the different sub-events can be
very close to each other in phase space).  The study of the effects of
non-flow correlations (from the point of view of a flow analysis) in
real data is rather complicated, so we start with the analysis of
Monte-Carlo events.

{\bf Non-flow correlations in Monte-Carlo generated events}.
When the true reaction plane is known (as it is in any generated
event) the contribution of non-flow correlations can be studied by
analyzing correlations along the axis perpendicular to the reaction
plane ($y$ axis).  For example, let us consider the correlation
between $\bQ^a$ and $\bQ^{b}$, the vectors (see Eq.(\ref{eqv}))
defined by two independent sub-events.  One can think of these vectors
as the total transverse momentum of all particles of the sub-event
(which is the case if the transverse momentum is used as a weight in
Eq.(\ref{eqv})).  Then if there are no other correlations except flow
\be
\la \bQ^a \bQ^{b} \ra =\la \bQ^a \ra \la \bQ^{b} \ra
= \la Q_x^a\ra \la Q_x^{b} \ra,
\ee
the very relation our first method is based on. It was assumed here
that the two $Q$-vectors are totally uncorrelated except that both of
them are correlated with the reaction plane.\footnote{Strictly
speaking there is one more assumption here, namely that the strength
of the flow does not fluctuate event by event.}  If this is not true
and there exist other correlations, then their contribution in first
order would be the same to the correlations of $x$ components and $y$
components.  Then
\bea
\la Q_x^a Q_x^{b} \ra &=& 
\la Q_x^a\ra \la Q_x^{b} \ra +
\la Q_x^a Q_x^{b} \ra_{non-flow} 
\nonumber \\ &\approx&  
\la Q_x^a\ra \la Q_x^{b} \ra +
\la Q_y^a Q_y^{b} \ra_{non-flow} 
=
\la Q_x^a\ra \la Q_x^{b} \ra +
\la Q_y^a Q_y^{b} \ra
\eea

\be
\la Q_x^a \ra \la Q_x^{b} \ra 
\approx  
\la Q_x^a Q_x^{b} \ra
-\la Q_y^a Q_y^{b} \ra.
\ee
Analyzing the correlation $\la Q_y^a Q_y^{b} \ra$ (and in particular $\la
\sin(n(\Psi_n^a-\Psi_r)) \sin(n(\Psi_n^{b}-\Psi_r))\ra$ one can
estimate the corrections to the formulae (8,12-15).

{\bf Non-flow correlations in real data}.
The direct application of the above described method to real data is
not possible. What can be done is the analysis of similar correlations
using, instead of $\Psi_r$, the event plane derived from the second
harmonic, where as the analysis of different models shows, the
contribution of non-flow effects is significantly less.  Then with the
(second harmonic) event plane resolution known one can carry out the
above analysis.
 
There exists in the literature other methods for estimating and
accounting for non-flow correlations, see, for
example refs.~\cite{dan88,olli95}. Here we briefly describe the
method~\cite{olli95}, which was applied to the data of the WA93
Collaboration~\cite{heer96}.  It was proposed~\cite{olli95,heer96} to
characterize the non-flow correlation contribution by the value of the
parameter
\be
c=
\frac{\la \bQ^a \bQ^{b} \ra - \la Q_x^a\ra \la Q_x^{b} \ra}
{\sqrt{\la (\bQ^a)^2 \ra }\sqrt{\la (\bQ^b)^2 \ra }}
\approx
\frac{\la \bQ^a \bQ^{b} \ra - \la Q_x^a\ra \la Q_x^{b} \ra}
{N}
,
\ee
where $N$ is the sub-event multiplicity and, as in section~\ref{sebe},
for simplicity we assume $w_i=1$.  The parameter $c$ can strongly
depend on the particular choice of sub-events, but if the non-flow
contribution is dominated by two-particle correlations, it is largely
independent of multiplicity~\cite{olli95}.  The non-flow correlation
changes the distribution of event flow vectors, and in particular
Eq.~(\ref{r2}) now reads
\be
\la Q_n^2 \ra \approx \la N \ra + \bar{v}_n^2 \la N^2 \ra + 2c \la N \ra,
\ee
If the parameter $c$ is relatively large, it can bias the results
derived from an application of Eq.~(\ref{r2}) to the data.  However,
the non-flow correlations contribute to the {\it shape} of the
distribution in $Q_n$ (Eq.~(\ref{ebess})) in a different way than
flow.  They produce mostly a change in $\sigma$, the ``Gaussian
width'' of the distribution (Eq.(\ref{sigma})), which is modified to
\be
\sigma^2=\frac{1 }{2N} (1+2c),
\ee
and the parameter $c$ can be directly extracted from the data.  The
analysis~\cite{heer96} of S+S and S+Au data gave (for their particular
sub-event selection) $c\approx0.034\pm0.025$ which is not negligible
compared with $v_2\approx0.04$--0.05 found in this analysis at
multiplicities of about 30--50.  Another possible consequence of
non-flow contributions could be the change in the shape of the
distribution of the difference of flow angles of
sub-events~\cite{olli95}.

\section{Simple way to introduce flow in a Monte-Carlo event generator}
Sometimes in order to investigate different detector effects or the
reliability of the method, one needs to introduce anisotropic flow
into a Monte-Carlo event generator.  It can be done by changing the
azimuthal angle of each particle (and consequently changing the
density in the azimuthal angle space)
\be
 \phi \rightarrow \phi'=\phi + \Delta \phi,
\label{phi}
\ee
where 
\be
 \Delta \phi = \sum_n \frac{-2}{n} \tilde{v_n} \sin(n(\phi - \psi_0)),
\label{dphi}
\ee
$\tilde{v_n}$ are the $n$ parameters of the transformation, and
$\psi_0$ is the direction of the added flow. The $\tilde{v_n}$ can be
functions of rapidity and transverse momentum and, in particular, the
$\tilde{v_n}$ for $n$ odd should reverse sign in the backward
hemisphere. One can check that such a change in the azimuthal angle
results in the required change in the distribution.  To first order in
$\tilde{v_n}$, $v_n \equiv \la \cos(n(\phi'-\Psi_r))
\ra =\tilde{v_n}$.  Small higher order corrections ($\sim v_n^2$), if needed,
can also be easily calculated, for example, numerically.

\section{Discussion}

{\bf Higher harmonics ($n\geq3$)}.
Note that the flow analysis methods presented in this paper are valid
for all harmonic orders of anisotropy. The higher harmonics look at
the event with higher symmetry on a finer scale. It should be
emphasized that the study of anisotropic flow corresponding to
harmonic orders of $n\geq 3$ has interesting aspects.  For instance,
one would expect large differences between the theoretical predictions
of hydro and cascade type models in the higher harmonics of the
particle azimuthal distributions. Also there are physics processes
which could lead to non-zero higher harmonics. It is widely accepted
that one of the main reasons for pion ``anti-flow'', directed flow in
the direction opposite to that of the protons, is caused by pion
shadowing by co-moving nucleons. Such shadowing could affect pion
higher harmonic azimuthal distributions, and affect them differently
at different rapidities.  Another effect is the transition from
out-of-plane elliptic flow (squeeze-out) which is very important at
low energies to in-plane elliptic flow which is the main effect at
high energies. At the transition beam energy both effects may be
important and the fourth harmonic may peak at this energy. Even at
high beam energies, the out-of-plane squeeze-out effect could dominate
at short times and the in-plane expansion at long times, leading to an
overall fourth order harmonic coefficient.

{\bf Transverse radial and anisotropic flow.}
We would like also to emphasize that the above methods permit the
reconstruction of the triple differential distribution with respect to
the reaction plane, and in particular are convenient for the analysis
of the $p_t$ dependence of the anisotropy.  The importance of such a
study was stressed in~\cite{lvrd}, where the effect of the interplay
of (transverse) radial and directed flow has been studied.  In that
analysis transverse directed flow is considered a result of the
movement of an effective source in the transverse plane.  It is
assumed that in the source rest frame the first moment of the
azimuthal distribution ($v_1$) is zero, and the final anisotropy
appears only as a consequence of the source movement in the transverse
direction.

In this case the transverse momentum dependence of $v_1$ has a rather
specific shape.  The radial expansion results in decreasing $v_1$ at
low $p_t$.  For some sets of parameters (temperature, radial and
directed velocities) it becomes negative.  Physically it corresponds
to the case, when particle production with such a value of $p_t$ is
more probable from the part of the effective source which moves in the
opposite direction from the flow direction, because in this case the
directed flow and the radial flow tend to compensate each other.

The same considerations can be applied to elliptic flow as well. If
one assumes that elliptic flow is a consequence of more rapid
expansion of the effective source in some plane, then the $p_t$
dependence of $v_2$ would exhibit exactly the same features as has
been observed for $v_1$ in the case of directed flow.

{\bf Pair-wise azimuthal correlations.}
Two-particle large-angle azimuthal correlations are often proposed as
a tool to study anisotropic flow~\cite{wang91}.  Not rejecting this
possibility we note that the expected signal in such correlations can
be very small in magnitude.  It can be easily shown that the pair-wise
distribution in the azimuthal angle difference ($\Delta \phi =\phi_1
-\phi_2$) is
\be
 \frac{dN^{pairs}}{d\Delta \phi} \propto (1 + 
\sum_{n=1}^\infty 2 v_n^2 \cos(n \Delta \phi)) .
\ee
Note that the ``signal'' is $v_n^2$, and thus is small. 
It does not imply though, that the ``signal to noise ratio'' is small,
and the method, in principle, can be successfully applied to data to
obtain the flow signal.  However the reconstruction of the triple
differential distribution with respect to the reaction plane (the goal
of the flow analysis) becomes more involved.

This method does not require the determination of the event
plane. Usually such a distribution is constructed using all possible
pair combinations in each event. Note that in this case each particle
enters into many pairs (on the order of the mean multiplicity of an
event) and, consequently, the pairs are not statistically independent.
Thus special precautions have to be taken for the evaluation of the
error of the results~\cite{lvmix}.

\section*{Acknowledgments}
The discussions with G.~Cooper, S.~Esumi, G.~Rai, H.-G.~Ritter, and
T.~Wienold as well as with many other members of the E877, NA49, and
E895 Collaborations are appreciatively acknowledged.  One of the
authors (S.V.)  thanks the Nuclear Science Division at LBNL for
financial support during his visit to LBNL, where this work was
started. This work was supported by the US Department of Energy under
Contract DE-ACO3-76SFOOO98.

\clearpage
\newpage

\end{document}